\documentclass{emulateapj}
\usepackage{graphicx}
\usepackage{epsfig}
\usepackage{rotating}

\shorttitle{X-ray Emission from the nearby PSR B1133+16 }
\shortauthors{Kargaltsev, O., Pavlov G.\ G., and Garmire G.\ P.}

\def\chan{{\sl Chandra}}
\def\xmm{{\sl XMM-Newton}}

\newcommand{\gapr}{\raisebox{-.6ex}{\mbox{
$\stackrel{>}{\mbox{\scriptsize$\sim$}}\:$}}}

\begin{document}

\title{
X-ray Emission from the Nearby PSR B1133+16 and Other Old Pulsars}

\author{
 O.\ Kargaltsev, G.\ G.\ Pavlov, and G.\ P.\ Garmire}
\affil{The Pennsylvania State University, 525 Davey Lab, University
Park, PA 16802, USA}
\email{green@astro.psu.edu,pavlov@astro.psu.edu}

\begin{abstract}
We detected a nearby ($d=360$ pc), old ($\tau = 5$ Myr) pulsar
B1133+16 with \chan. The observed pulsar's flux is $(0.8\pm
0.2)\times 10^{-14}$ ergs cm$^{-2}$ s$^{-1}$ in the 0.5--8 keV band.
Because of the small number of counts detected, the spectrum can be
described by various models. A power-law fit of the spectrum gives a
photon index $\Gamma \approx 2.5$ and an isotropic luminosity of
$1.4\times 10^{29}$ ergs s$^{-1}$ in the 0.5--8 keV band, which is
about $1.6\times 10^{-3}$ of the spin-down power $\dot{E}$. The
spectrum can also be fitted by a blackbody model with a temperature
of $\approx 2.8$ MK and a projected emitting area of $\sim 500$
m$^2$, possibly a hot polar cap. The X-ray properties of PSR
B1133+16 are similar to those of other old pulsars observed in
X-rays, particularly the drifting pulsar B0943+10.

\end{abstract}
\section{Introduction}

Models of neutron star (NS) cooling (e.g., Yakovlev \& Pethick 2004)
predict that at an age of $\gapr 1$ Myr at least passively cooling
NSs become too cold to emit X-rays from the bulk of NS surface.
Therefore, X-ray emission from isolated radio pulsars of such old
ages is expected to consist of a magnetospheric component and,
possibly, a thermal component emitted from small areas (polar caps
[PCs]) heated by relativistic particles created in the pulsar's
acceleration zones (e.g., Harding \& Muslimov 2001, 2002; hereafter
HM01 and HM02). Hence, studying the X-ray emission from old pulsars
allows one to examine the properties and evolution of magnetospheric
radiation, probe the particle acceleration mechanisms operating in
the magnetospheres, and constrain the PC heating and emission
models.

Although most of the $\sim 1600$ currently known radio pulsars are
older than 1 Myr (see Fig.\ 1), they are intrinsically faint X-ray
sources because the luminosities of both the magnetospheric and PC
components are fractions of the spin-down power $\dot{E}$, which
decreases with pulsar's age. So far, only seven pulsars with
characteristic ages $\tau \equiv P/(2\dot{P}) > 1\,{\rm Myr}$ have
been detected in X-rays\footnote{The oldest known pulsars are the
recycled millisecond pulsars (MSPs) with ages 0.1--10 Gyr. Since
they are spun-up by accretion in binaries, their properties may
differ from those of the ``ordinary'' old isolated pulsars;
therefore, we do not consider MSPs in this paper.} (i.e., about 10\%
of X-ray detected ``ordinary'' pulsars), all of them at distances
$<$2 kpc (Zavlin \& Pavlov 2004 [hereafter ZP04];
Becker et al.\ 2004,2005; Zhang et
al.\ 2005; \"{O}gelman
\& Tepedelenlio\v{g}lu 2005).

In principle, the magnetospheric and PC components of X-ray
radiation can be distinguished by their spectra and pulse shapes
(e.g., the magnetospheric radiation is expected to have a harder
spectrum and show sharper pulsations). However, even the brightest
of the detected old pulsars are too faint to establish the spectral
shape unambiguously. For instance, the spectrum of the relatively
bright PSR B0950+08 ($d=260$ pc, $\tau = 17$ Myr, $F_{0.2-10\,{\rm
keV}} = 1.1\times 10^{-13}$ ergs s$^{-1}$ cm$^{-2}$), observed
recently with \xmm, can be fitted with either a single power-law
(PL) model with photon index $\Gamma \approx 1.75$ or a combination
of a PL model ($\Gamma \approx 1.35$)
 and a thermal (hydrogen atmosphere) model, with PC temperature
$T_{\rm pc} \sim 1$ MK and radius $R_{\rm pc}\sim 250$ m (the
two-component interpretation is supported by the energy-dependent
pulse shape; ZP04). Therefore, it is important to observe a larger
sample of old pulsars and study their properties via comparative
analysis.

In this {\em Letter}, we report on first X-ray detection of one of the
nearest pulsars, PSR B1133+16 (= J1136+1551; see Table 1 for the
radio pulsar parameters). This old pulsar ($\tau = 5$ Myr) has the
largest proper motion, 375 mas yr$^{-1}$, among the known radio
pulsars, which corresponds to the transverse velocity $V_\perp
\simeq 630$ km s$^{-1}$ at the distance of 360 pc inferred from the
radio parallax measurement (Brisken et al.\ 2002). The pulsar shows
a double-peaked radio pulse, and it is known to spend $\simeq15\%$
of the time in a ``null state'' where it does not emit radio pulses
(Biggs 1992). From radio polarization observations, Lyne \&
Manchester (1988) infer the angle $\alpha = 51.3^\circ$ between the
magnetic and rotation axes, and the angle $\beta = 3.7^\circ$ of the
closest approach of the magnetic axis to the line of sight. Although
the pulsar's spin-down power, $\dot{E} = 8.8\times 10^{31}$ ergs
s$^{-1}$, is smaller than those of the other pulsars detected in
X-rays, its ``spin-down flux'', $\dot{E}/(4\pi d^2) = 5.8\times
10^{-12}$ erg s$^{-1}$, is large enough to warrant an exploratory
X-ray observation with \chan\ or \xmm. We describe a {\sl Chandra}
observation of PSR B1133+16
 and its analysis in \S2 and \S3
and compare the properties of this pulsar with those of other old
pulsars in \S4.

\section{Observations}

\begin{deluxetable}{ll}
\tabletypesize{\scriptsize} \tablewidth{9cm}
\setlength{\tabcolsep}{0.001in} \tablecaption{ Observed and derived
parameters for PSR B1133+16} \tablehead{\colhead{Parameter} &
\colhead{Value} } \startdata
R.A.~(J2000)...........................................
& ~~~~$11^{\rm h}36^{\rm m}03\fs1829(10)$ \\
Dec.~(J2000)...........................................
& ~~~~$+15^\circ 51' 09\farcs726(15)$ \\
Epoch of position (MJD)........................
& ~~~~51,544.0 \\
Proper motion, R.A./Dec.\ (mas yr$^{-1}$)..
& ~~~~$-$73.95(38)/368.05(28) \\
Spin period, $P$~(s).................................
&  ~~~~1.1879 \\
Dispersion Measure, DM (cm$^{-3}$~pc)\,.......
& ~~~~4.86 \\
Distance from parallax, $d$ (pc)................
& ~~~~357(19) \\
Surface magnetic field, $B_s$~($10^{12}$ G).........
& ~~~~2.13\\
Spin-down power, $\dot{E}$~($10^{31}$ erg s$^{-1}$).......
& ~~~~8.8 \\
Age, $\tau=P/(2\dot{P})$, (Myr).........................
& ~~~~5.04 \\
\enddata
\tablecomments{Based on the data from the ATNF radio pulsar catalog
(Manchester et al.\ 2005). The position, proper motion, and distance
are from Brisken et al.\ (2002). The numbers in parentheses
represent uncertainties in the last significant digits. }
\end{deluxetable}

PSR B1133+16 (hereafter B1133) was observed with the Advanced CCD
Imaging Spectrometer (ACIS) on 2005 February 23 (start time
53,424.79 MJD).
 The useful scientific exposure time was 17,911 s. The
observation was carried out in Very Faint mode, and the pulsar was
imaged at the aim point on ACIS-S3 chip. The detector was operated
in full frame mode which provides time resolution of 3.2 s. The data
were reduced using the Chandra Interactive Analysis of Observations
(CIAO) software (ver.\ 3.2.1; CALDB ver.\ 3.0.3).

\section{X-ray image and spectrum}

Figure 2 shows the ACIS-S3 image of the B1133 field. The X-ray source
 is clearly seen
at R.A.$=11^{\rm h}36^{\rm m}03\fs169$, Dec.$=+15^{\circ}51'
11\farcs87$ (centroid uncertainty is 0\farcs2). This position is
within 0\farcs1 of the radio position of B1133 derived from its
proper motion at the epoch of the {\sl Chandra} observation.
Therefore, we conclude with confidence that we detected the pulsar's
X-ray counterpart. The distribution of source counts in the image is
consistent with that of a point source.

\begin{figure}
 \centering
 \hspace{-1.7cm}
\includegraphics[width=2.9in,angle=90]{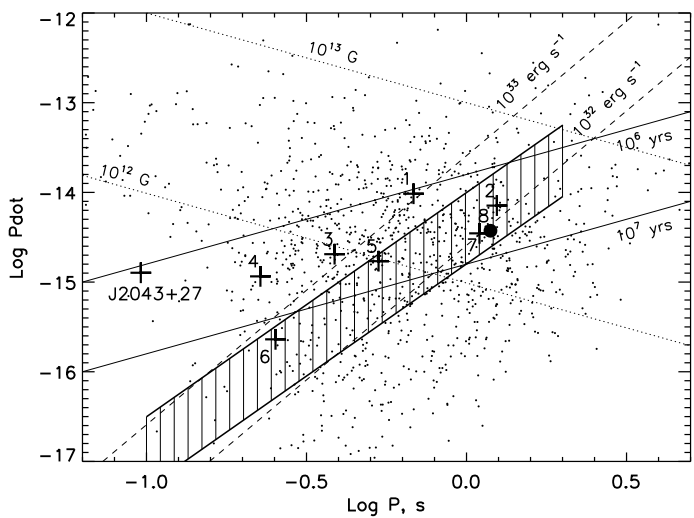}
\caption{ $P$-$\dot{P}$ diagram for $\approx1400$ radio pulsars
(dots) from the ATNF catalog (Manchester et al.\ 2005). Lines of
constant pulsar age, magnetic field, and $\dot{E}$ are shown. Eight
pulsars with $\tau>1$ Myr that have been previously observed in
X-rays are marked by crosses, and PSR B1133+16 is marked by the
filled circle. The numbers, 1 through 8, near the marked pulsars
correspond to those in Fig.\ 5. The hatched area represents
plausible locations of the death line for the curvature radiation
 induced cascade (from eqn.\ [52] of HM02, for pair production
 efficiencies in the $0.2-0.5$ range.) }
\end{figure}

We extracted the pulsar's spectrum from a circular aperture with
the radius of 2.5 ACIS pixels ($=1\farcs23$; $\approx90$\% encircled energy radius)
 using the CIAO {\em psextract} task. The background
was extracted from $5''<r<22''$ annulus centered on the source. The
total number of counts within the source aperture is 33, of which
99.7\% are expected to come from the source. The observed source
flux is $(0.8\pm 0.2) \times 10^{-14}$ ergs cm$^{-2}$ s$^{-1}$ in
the 0.5--8 keV band. We group the counts into 7 spectral bins (4--5
counts per bin; see Fig.\ 3) and fit the spectrum with absorbed PL
and blackbody (BB) models. Because of the small number of counts, we
have to freeze the hydrogen column density, $n_{\rm H}$.
 We adopt $n_{\rm H} =
1.5\times 10^{20}$ cm$^{-2}$ estimated from the dispersion measure,
4.86 cm$^{-3}$ pc, assuming a 10\% degree of ionization of the ISM.
(Varying $n_{\rm H}$ in the plausible range of 1--$2\times 10^{20}$
cm$^{-2}$ does not change the fits substantially.) Since each of
the spectral bins has only a few counts, we fit the
spectrum using the maximum likelihood
method implemented in XSPEC (v.\ 11.3) with the C-statistic (Cash 1979).
The fitting parameters for the PL and BB models
 are given in Table 2 while
the fits and the confidence contours are shown in Figures 3 and 4.

\begin{figure}[t]
 \centering
\includegraphics[width=2.6in,angle=0]{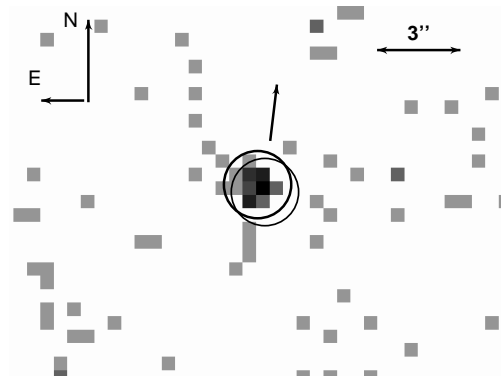}
\caption{ ACIS-S3 image of PSR B1133+16. The two circles are
centered at the centroid of the source count distribution and the
pulsar's position predicted from its proper motion. The arrow shows
the direction of the proper motion; its length corresponds to the
pulsar's displacement during 6 years. }
\end{figure}

The PL fit gives a photon index $\Gamma = 2.2$--2.9
 and an isotropic
luminosity $L_{0.5-8\,\rm keV}=4\pi d^2 F_{0.5-8\,\rm keV}^{\rm
 unabs} \approx 1.4\pm0.3\times 10^{29}$
 ergs s$^{-1}$, for $d=357$ pc.
The temperature and the projected area of the emitting region
obtained from the BB fit are strongly correlated (see Fig.\ 4),
which results in large uncertainty of these parameters. However, the
projected area, $\sim 500$ m$^2$,
 is much smaller than that
of a NS, $\sim 3\times 10^{10}$ m$^2$, which suggests that the
radiation could be emitted from hot polar caps, with an effective
radius $R \sim 13 \langle \cos\theta\rangle^{-1/2}$ m
 and a
bolometric luminosity $L_{\rm bol}= \sigma T^4 A \langle
\cos\theta\rangle^{-1}= 3.2^{+0.5}_{-0.6} \times 10^{28}\langle
\cos\theta\rangle^{-1}$ ergs s$^{-1}$,
 where $\langle
\cos\theta\rangle$ is a time-averaged cosine of the angle between
the line of sight and the magnetic axis ($\approx 0.47$
 for the
axis orientations suggested by Lyne \& Manchester 1988).

\begin{figure}
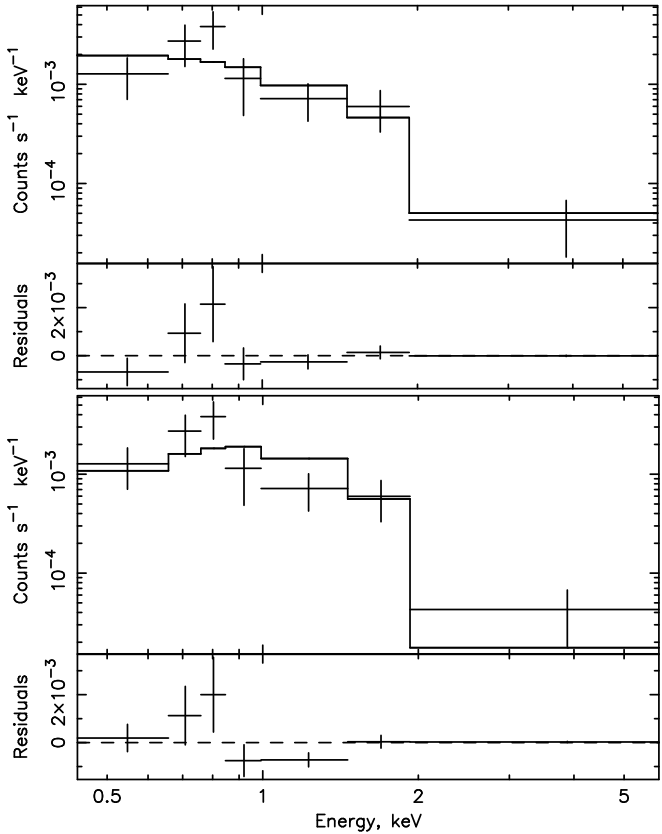

 \centering
 \vbox{
\includegraphics[width=2.03in,angle=-90]{f3a.eps}
\includegraphics[width=2.3in,angle=-90]{f3b.eps}}
\caption{ B1133 spectrum fitted with the PL (top) and BB (bottom)
models. The contributions of the energy bins into the best-fit
C-statistic are shown, multiplied by $-$1 when the number of data
counts is smaller than the number of model counts. }
\end{figure}

\begin{figure}[bp]
 \centering
 \vbox{
\includegraphics[width=2.5in,angle=90]{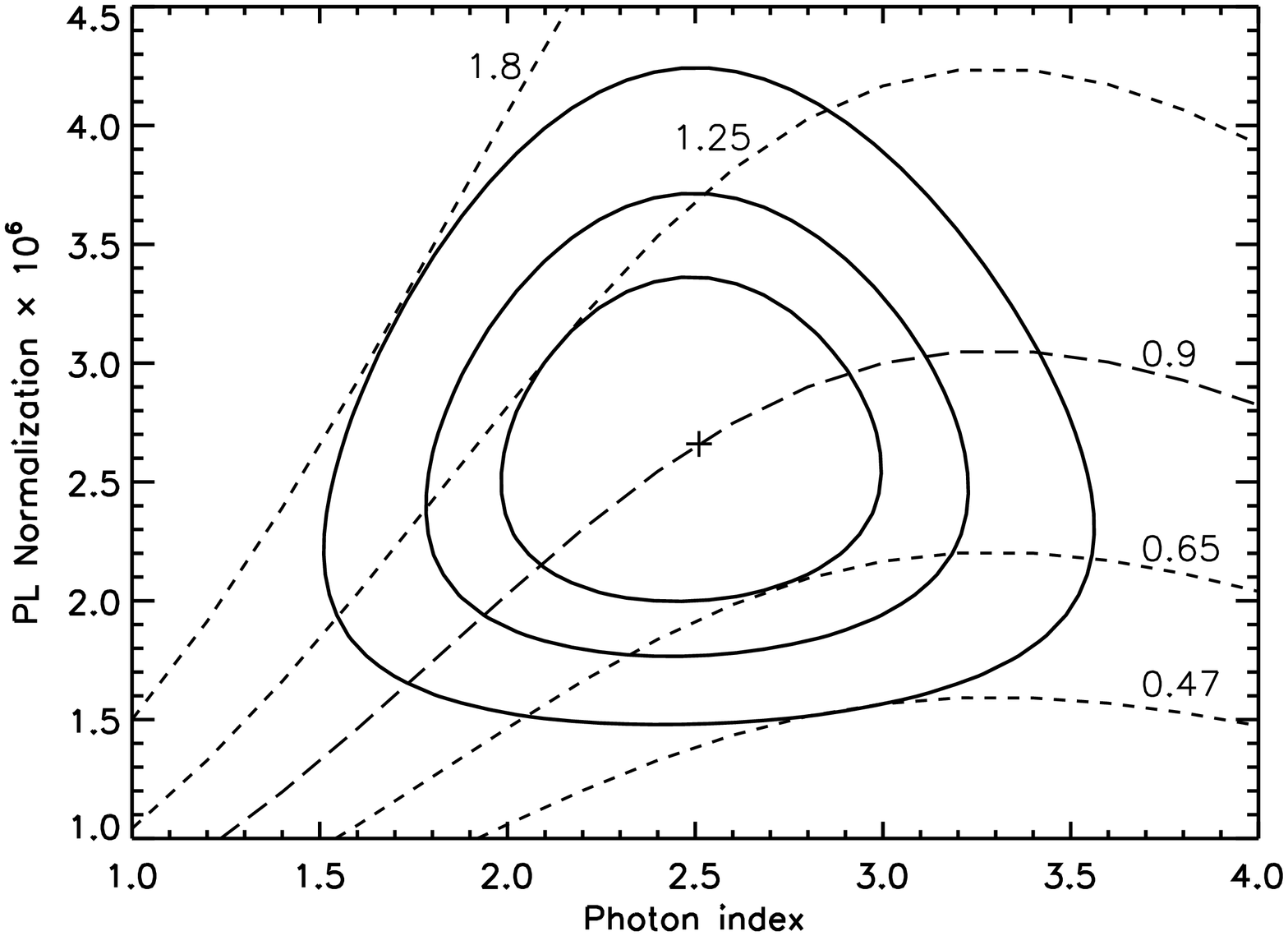}
\vspace{-0.0cm}
\includegraphics[width=2.5in,angle=90]{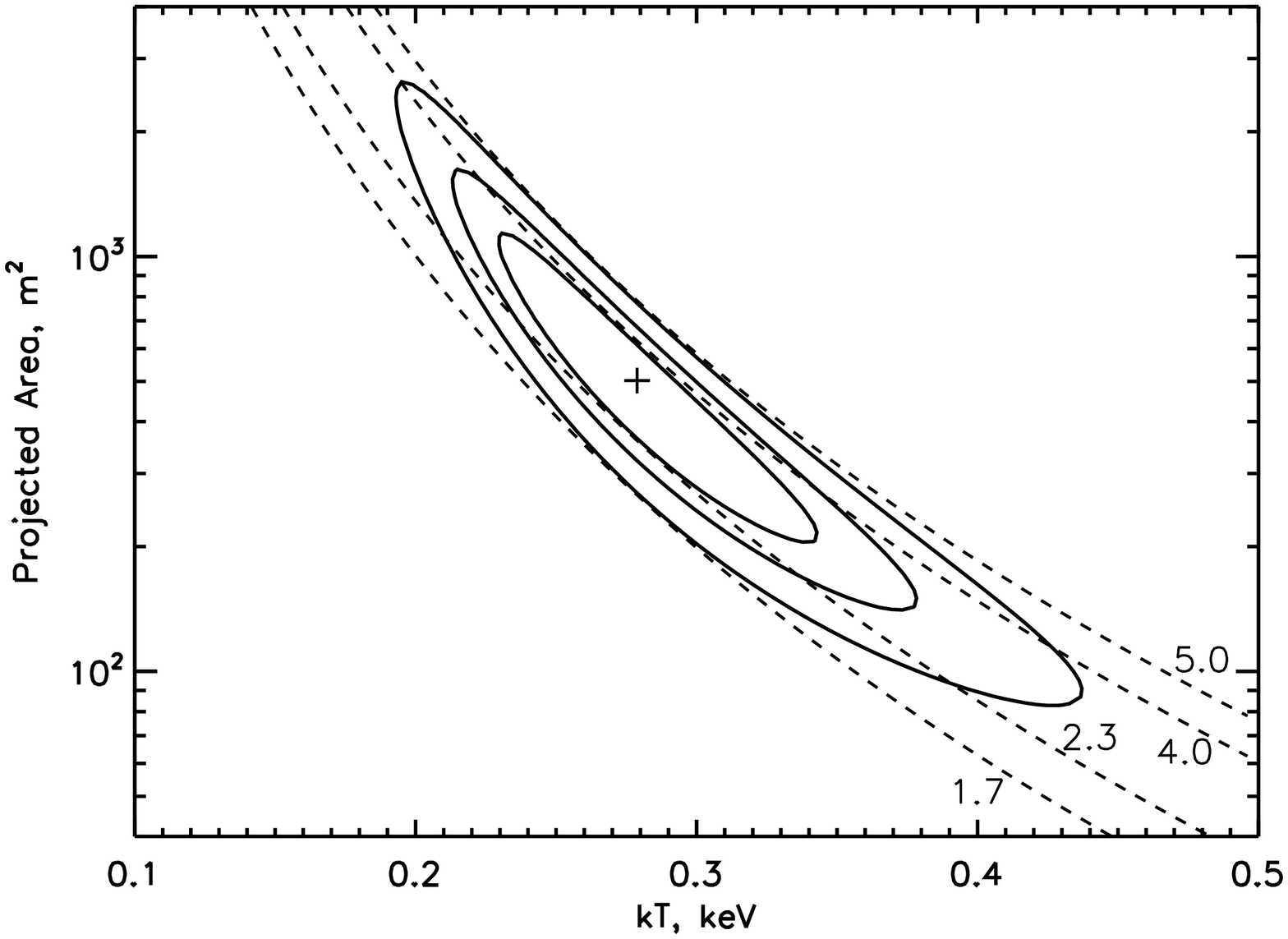}}
\caption{Confidence contours (68\%, 90\%, and 99\%, computed for two
interesting parameters) for the PL (top) and BB (bottom) model fits
to the ACIS spectrum of PSR B1133+16. The PL normalization is in
units of $10^{-6}$ photons cm$^{-2}$ s$^{-1}$ keV$^{-1}$ at 1 keV.
The BB normalization (vertical axis) is the projected emitting area
in units of m$^2$, for d = 357 pc. The lines of constant unabsorbed
flux (PL model; in units of $10^{-14}$ erg cm$^{-2}$ s$^{-1}$) and
constant bolometric luminosity (BB model; for $\langle
\cos\theta\rangle =1$, in units of $10^{28}$ erg s$^{-1}$) are
plotted as dashed lines, for fixed $n_{\rm H} =1.5\times 10^{20}$
cm$^{-2}$.  }
\end{figure}

\section{Discussion.}

\begin{table}[tbp]
 \caption{Fitting
parameters for the PL and BB models} \vspace{-0.5cm}
\begin{center}
\begin{tabular}{cccccc}
\tableline\tableline Model & Norm. & $\Gamma$ or $kT$ &
 $C$/dof & $Q$ & $L_{29}$ \\
\tableline
 PL   & $2.66^{+0.43}_{-0.41}$ & $2.51^{+0.36}_{-0.33}$ & 6.20/5 &43\% & $1.4\pm0.3$ \\
 BB   & $0.50^{+0.30}_{-0.22}$ & $0.28^{+0.04}_{-0.03}$ & 10.4/5 &79\% & $0.32^{+0.05}_{-0.06}$ \\
     \tableline
\end{tabular}
\end{center}
\tablecomments{The hydrogen column density was fixed at $n_{\rm H} =
1.5\times 10^{20}$ cm$^{-2}$ for both the PL and BB fits.
Normalization in second column is the spectral flux in $10^{-6}$
photons cm$^{-2}$ s$^{-1}$ keV$^{-1}$ at 1 keV for the PL fit, and
the projected area in $10^7$ cm$^2$ for the BB fit. Third column
gives the photon index for the PL fit and the temperature in keV for
the BB fit.
Fourth column gives the best-fit value of C-statistic and the number
of degrees of freedom. The parameter $Q$ in fifth column is the
percentage of 10,000 Monte Carlo simulations, drawn from the
best-fit model, which give a C-statistic value lower than the
best-fit value.
The last column gives the luminosities in units of $10^{29}$ ergs
s$^{-1}$, for $d=357$ pc ($L_{\rm 0.5-8\, keV}$ for the PL fit,
$L_{\rm bol}\langle\cos\theta\rangle$ for the BB fit,
see text).}
\end{table}

The scarce statistics of the short observation of B1133 does not
allow one to
 differentiate between the
alternative spectral models.
 Below we
explore various possibilities by invoking independent arguments such
as comparison with the properties of other old pulsars observed in
X-rays.

\begin{figure}[t]
\vbox{
 \centering
\includegraphics[width=2.6in,angle=90]{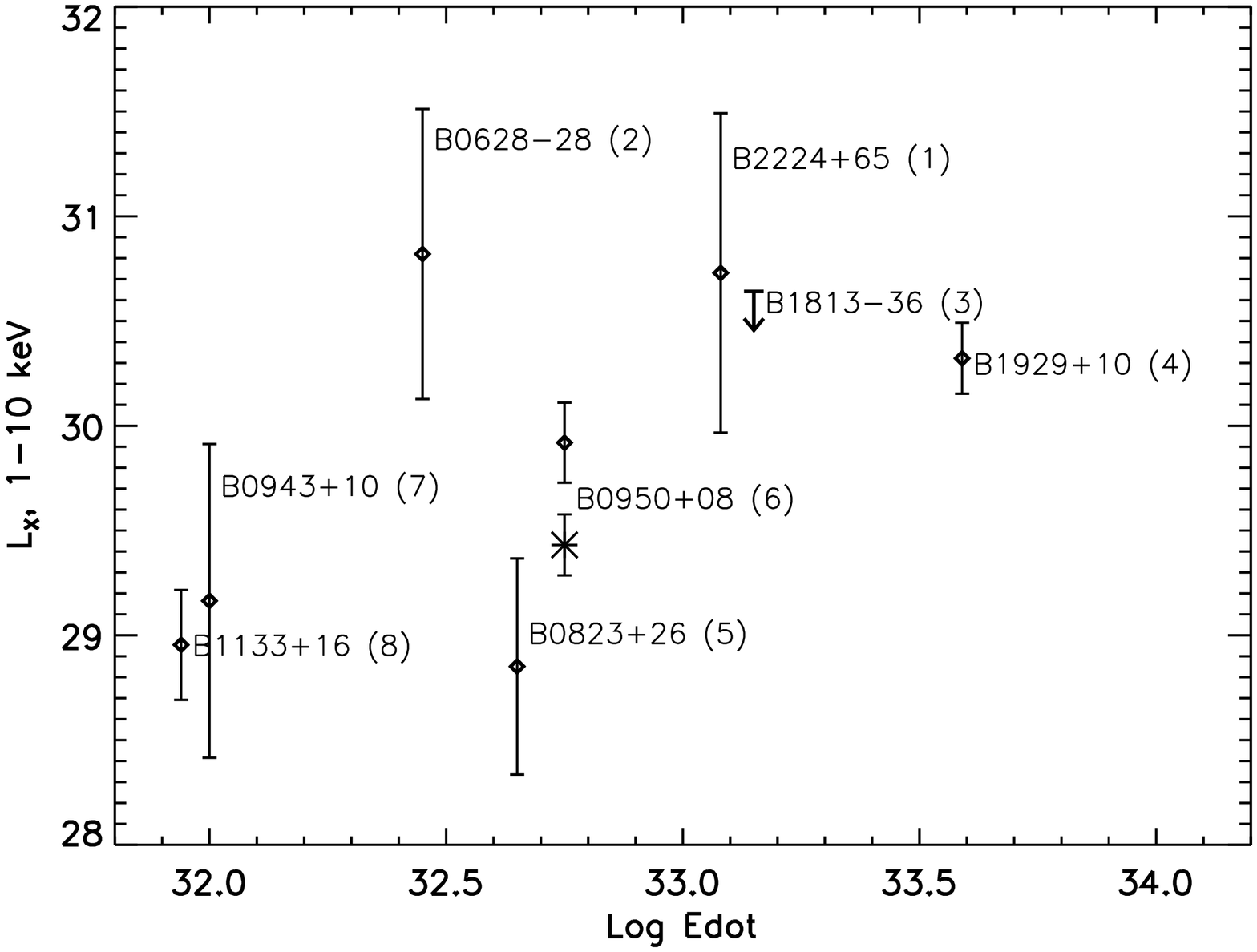}
\vspace{-0.0cm}
\includegraphics[width=2.6in,angle=90]{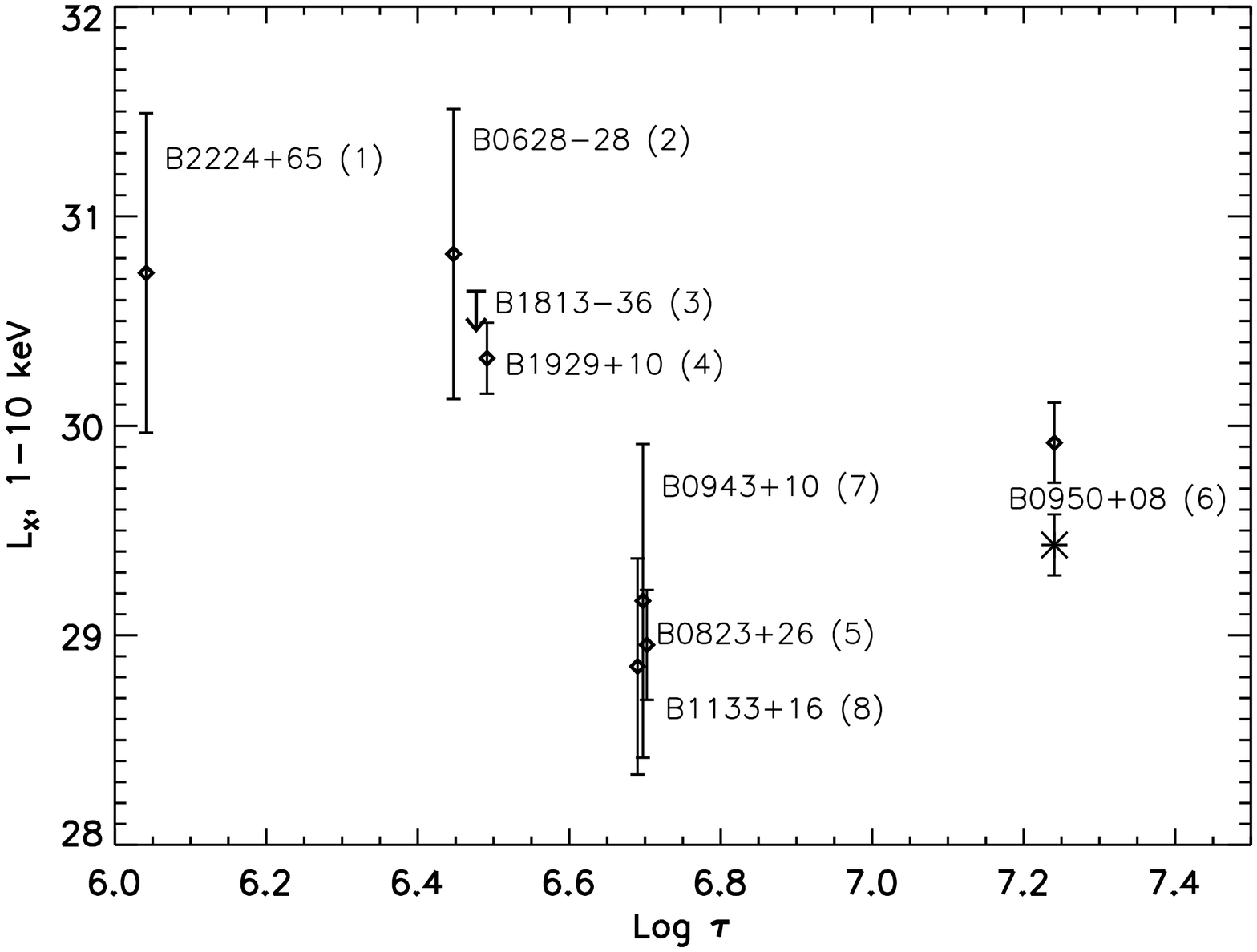}
} \caption{The X-ray luminosities (1$-$10 keV band) of eight old
pulsars
 versus
spin-down power ({\sl top}) and characteristic age ({\sl bottom}).
The numbers in parentheses correspond to the pulsars marked in Fig.\
1. For PSR B0950+08, the diamond and the asterisk show the
luminosities of the nontermal and and thermal components,
respectively (as determined from the PL+NSA fit by ZP04). For the
other pulsars, the luminosities were obtained from PL fits found in
the literature (ZP04; \"{O}gelman, H., \& Tepedelenlio\v{g}lu 2005;
Becker et al.\ 2004, 2005; Zhang et al.\ 2005) or remeasured from
the {\sl Chandra} data. PSR B1813--36 was not detected in a 30 ks
{\sl Chandra} exposure, hence only the upper limit is plotted. }
\end{figure}

For the PL model, the photon index $\Gamma \approx 2.5$
 of the B1133
spectrum is somewhat  larger than $\Gamma\simeq 1$--2 observed for
young and middle-aged pulsars (e.g., Gotthelf 2003). Interestingly,
PL fits of some other old pulsars (e.g., B0943+10, B0628$-$28) also
show very soft spectra, with $\Gamma \approx 2$--3 (ZP04; Zhang et
al.\ 2005; \"{O}gelman, \& Tepedelenlio\v{g}lu 2005), suggesting
that pulsar spectra might soften with increasing age or decreasing
spin-down power. However, the example of PSR B0950+08, the oldest in
the sample, demonstrates that the softness of the one-component PL
fit may be caused by the presence of a thermal (PC) component, which
mostly contributes at lower energies: a single PL fit gives $\Gamma
\approx 1.8$ for PSR B0950+08, while the slope of the PL component
in the two-component fit is $\approx$1.3, similar to younger pulsars
(ZP04).

The extrapolation of the X-ray PL fit into the optical range
predicts easily detectable optical magnitudes (e.g., $V\sim 20$--24
mag). A more realistic prediction, $V\sim 29$--30 mag, can be
obtained from the empirical relation between the optical and X-ray
fluxes of spin-powered pulsars (ZP04).

The luminosity found from the PL fit allows one to estimate the
``X-ray efficiency'' for B1133: $\eta \equiv L_X/\dot{E} =
1.6(-0.4,+0.3)\times 10^{-3}$ in the 0.5--8 keV band ($\eta = 1.0\pm
0.3\times 10^{-3}$ and $0.6\pm 0.2\times 10^{-3}$ in the 1--10 and
2--10 keV bands, respectively). The inferred efficiency is larger
than for most young and middle-aged pulsars (e.g., Possenti et al.\
2002), but it is within the range of efficiencies found for other
old pulsars (ZP04). We plot in Figure 5 the 1--10 keV luminosities
of eight old pulsars, as inferred from the PL fits, versus spin-down
power and characteristic age\footnote{We exclude PSR J2043+2740 from
this sample because its predominantly thermal X-ray spectrum
resembles those of middle-aged pulsars rather than of old ones
(ZP04). Possibly, its characteristic age, $\tau = 1.2$ Myr, is
larger than its actual age. Notice that its position on the
$P$-$\dot{P}$ diagram is also quite different from those of the
other pulsars in the sample (see Fig.\ 1).}. For the pulsars whose
parallaxes have not been measured,  we ascribe a factor of 2
uncertainty to the distances estimated from the pulsar's dispersion
measure. Although the luminosity shows some correlation with
$\dot{E}$, the scatter is quite substantial (for instance, the
luminosities of B0628$-$28 and B0823+26 differ by about 2 orders of
magnitude despite the close values of $\dot{E}$). Such a scatter can
possibly be attributed to different
 orientations of the pulsar beams with respect to the observer.
Overall, the apparent X-ray efficiencies of the observed old pulsars
are somewhat higher than those of younger pulsars, but this can be
caused by the selection effect (the sample is biased in favor of
brighter objects). The correlation of the luminosity with the
characteristic age is even less pronounced; for instance, the
luminosity of the oldest PSR B0950+08 ($\tau = 17$ Myr) is a factor
of 10 higher than those of PSRs B1133+16, B0943+10, and B0823+26
($\tau \approx 5$ Myr). We should remember, however, that the
characteristic age can be substantially different from the true age,
so the dependence $L_X(\tau)$ may not accurately characterize the
pulsar evolution.

The best-fit temperature, $\approx 2.8$ MK, obtained from the BB
fit to the spectrum of B1133, is close to $\approx 3.1$ MK and
$\approx 1.8$ MK obtained from the BB fit for PSR B0943+10 and BB+PL
fit for PSR B0950+08, respectively. Although the effective radius of
the emitting region, $R_{\rm BB}\sim 19$ m,
 is highly
uncertain because of poor statistics, it is certainly smaller than
the conventional PC radius assuming a dipole magnetic field, $R_{\rm
pc}=(2\pi R^{3}/cP)^{1/2}\simeq 130$ m. A similar discrepancy is
also seen in the above-mentioned fits to the spectra of B0943+10
($R_{\rm BB}\sim 20$ m, $R_{\rm pc} \simeq 140$ m) and B0950+08
($R_{\rm BB}\sim 50$ m, $R_{\rm pc} \simeq 290$ m). Zhang et al.
(2005) suggest that the small X-ray emitting  area can be explained
assuming that only a small fraction of the PC area is heated by the
inflowing relativistic particles created in ``spark discharges''
above the PC (Ruderman \& Sutherland 1975). An independent support
of this interpretation is provided by the subpulse drifting observed
in B0943+10. No subpulse drifting has been reported for
 B1133+16, but this may be the
result of a different orientation of the pulsar beam. The
discrepancy can also be caused by a PC thermal spectrum being
different from the blackbody. For instance, fits with the magnetic
hydrogen atmosphere (NSA) models (Pavlov et al.\ 1995) usually give
a factor of 2 lower temperatures and a factor of 5--10 larger radii,
with about the same observed luminosity. ZP04 explored this
possibility for PSR B0950+08 and obtained a temperature of
$\approx$1.1 MK and an emitting radius of 250 m, close to the
$R_{\rm pc}=290$ m. We do not attempt to fit the spectrum of B1133
with the NSA models because too few counts were detected and, with
the time resolution of 3.2 s, we cannot extract the light curves of
the 1.19 s pulsations needed for an accurate analysis of the highly
anisotropic atmosphere radiation.

 As one can see from Table 2 and Figure 3, the quality of the BB
fit is worse than that of the PL fit because of an excess of
observed counts at $E>1.5$ keV. This indicates that the spectrum
might have both thermal and nonthermal (PL) components,
  similar to those
seen in many younger pulsars and, likely, in the old B0950+08
(Pavlov et al.\ 2002; ZP04). It is very likely that the radiation of
the other old pulsars, from which too few counts have been detected
to firmly establish the origin of their X-ray emission, also
consists of both thermal and nonthermal components. If this is the
case, the luminosities inferred from the BB fits should be
considered as upper limits. The current upper limits on bolometric
BB luminosities are close to the 1--10 keV PL luminosities shown in
Figure 5.

If the X-ray radiation from B1133 is dominated by the thermal
emission from PCs, then its PC efficiency is $\eta_{\rm pc}\equiv
L_{\rm bol}/\dot{E}=3.6_{-0.7}^{+0.6}\times 10^{-4} \langle \cos
\theta \rangle^{-1}$ ($\sim 8\times 10^{-4}$ for $\langle \cos
\theta \rangle =0.47$).
 This efficiency is two orders of
magnitude
 lower than those predicted
  by HM01 models
  for the curvature radiation (CR) induced pair cascade (Fig.\ 7 of HM01),
which might indicate that the CR cascade does not operate in B1133.
Indeed, this pulsar is very close to the ``CR death line''
in the $P$-$\dot{P}$ diagram (see Fig.\ 1), perhaps even below it,
 depending on exact value of pair production efficiency (see HM01 for details).
  Note, however, that the PC luminosities for the CR cascade derived
  by  Arons (1981) under somewhat different assumptions
  underestimate the observed $\eta_{\rm pc}$ by a factor of 100.
The PC efficiencies
 expected from the less efficient inverse Compton scattering (ICS) cascade,
which may play some role below the CR death line (see HM02), are aso 1$-$2
orders of magnitude
  lower than our observational estimate.
Therefore, if B1133 lies below the CR death line, then the observed
emission is most likely produced in the magnetosphere.

To summarize, the X-ray spectrum of B1133 can be described by either
a nonthermal PL model or a thermal BB model, or a combination of
those. The X-ray parameters of B1133 are particularly close to those
of PSR B0943+10, which is almost a twin of B1133+16 in terms of $P$
and $\dot{P}$ also. The nonthermal X-ray efficiencies of B1133 and
other old pulsars generally exceed those of younger pulsars, but it
is currently unclear whether this is a genuine property or it is
caused by selection effects. PL fits of old pulsars show, on
average, steeper spectra, but this may be caused by the presence of
a thermal component.
If the predictions of the current pair cascade models are correct,
then either the CR cascade is still operating in B1133 and X-rays
are emitted from the heated NS surface, or the bulk of X-ray emission
is produced in the pulsar's magnetosphere.
Since
B1133 is a typical representative of the majority of radio pulsars,
it would be important to better constrain its X-ray properties in a
deeper observation.

\acknowledgements
This work was partially supported by NASA grants NAG5-10865 and
NAS8-01128 and {\sl Chandra} award SV4-74018.


\begin{thebibliography}{}
\bibitem[]{386} Arons, J. 1981, ApJ, 248, 1099

\bibitem[]{387} Becker, W., Weisskopf, M.\ C., Tennant, A.\ F., Jessner, A.,
Dyks, J., Harding, A.\ K., Zhang, S.\ N. 2004, ApJ, 615, 908

\bibitem[]{388} Becker, W., Kramer, M.\ C., Jessner, A., et al. 2005, submitted to ApJ,
astro-ph/0506545

\bibitem[]{391} Biggs, J.\ D. 1992, ApJ, 394, 574

\bibitem[]{393} Brisken, W.\ F., Benson, J.\ M., Gross, W.\ M., \& Thorsett, S.\ E.
2002, ApJ, 571, 906

\bibitem[]{396}  Harding, A.\ K., \& Muslimov, A.\ G. 2001, ApJ, 556, 987 (HM01)

\bibitem[]{398} Harding, A.\ K., \& Muslimov, A.\ G. 2002, ApJ, 568, 862 (HM02)

\bibitem[]{399} Cash, W. 1979, ApJ, 228, 939

\bibitem[]{400} Gotthelf, E.\ V.\ 2003, ApJ, 591, 361

\bibitem[]{402} Lyne, A.\ G., \& Manchester, R.\ N.  1988, MNRAS, 234,
477

\bibitem[]{405} Manchester, R.\ N., et al. 2005, AJ, 129, 1993

\bibitem[]{406} \"{O}gelman, H., \& Tepedelenlio\v{g}lu, E. 2005,
ApJ Lett., in press (astro-ph/0505461)

\bibitem[]{407} Pavlov, G.\ G., Shibanov, Yu.\ A., Zavlin, V.\ E., \& Meyer, R.
1995, in The Lives of the Neutron Stars, ed.\ M.\ A.\ Alpar, \"{U}.\
Kizilo\u{g}lu \& J.\ van Paradijs (Dordrecht: Kluwer), 71

\bibitem[]{411} Pavlov, G.\ G., Zavlin, V.\ E., \& Sanwal, D. 2002, in
Proc.\ 270 Heraeus Seminar on Neutron Stars, Pulsars, and Supernova
Remnants, ed.\ W.\ Becker, H.\ Lesch, \& J.\ Tr\"umper (MPE Rep.\ 278;
Garching: MPE), 283

\bibitem[]{416} Possenti, A., Cerutti, R., Colpi, M., \& Mereghetti, S.
2002, A\&A, 387, 993

\bibitem[]{419} Ruderman, M., \& Sutherland, P.\ G. 1975, ApJ, 196, 51

\bibitem[]{421} Yakovlev, D.G., \& Pethick, C.\ J. 2004, ARA\&A, 42, 169

\bibitem[]{423} Zavlin, V.\ E., \& Pavlov, G.\ G. 2004, ApJ, 616, 452 (ZP04)

\bibitem[]{425} Zhang, B., Sanwal, D., \& Pavlov, G.\ G. 2005, ApJ, 624, L109

\end{thebibliography}
\end{document}